\documentstyle[aps,pre]{revtex}      
 
\begin {document}
 
\title {Dynamics of two interacting particles in classical billiards}
 
\author {Lilia Meza-Montes$^a$ and Sergio E. Ulloa}
\address {Department of Physics and Astronomy and Condensed Matter and Surface 
Sciences Program, \\ Ohio University, Athens, OH 45701-2979}
\address {$^a$Instituto de F\'{\i}sica UAP, Apdo.\ Postal J-48, Puebla, Pue. 
M\'exico 72080}
  
\date{8 March 1997} 
\twocolumn
\maketitle
\begin {abstract} 
 The problem of two interacting particles moving in a $d$-dimensional
billiard is considered here. A suitable coordinate transformation
leads to the problem of a particle in an unconventional hyperbilliard.
A dynamical map can be readily constructed for this general system,
which greatly simplifies calculations.  As a particular example, we
consider two identical particles interacting through a screened
Coulomb potential in a one-dimensional billiard. We find that the
screening plays an important role in the dynamical behavior of the
system and only in the limit of vanishing screening length can the
particles be considered as bouncing balls. For more general screening
and energy values, the system presents strong non-integrability with
resonant islands of stability.  \end{abstract}
 
\narrowtext

A system of two interacting bodies moving in an otherwise free space,
is one of the few integrable problems known.  The reduction to the
one-body central force problem allows a solution by
quadratures \cite{Gutzwiller}. However, once the translational
symmetries are broken, as when the system is placed inside a billiard,
the center-of-mass (CM) and angular momenta are in general no longer
constants of motion. In this case, the classical dynamics of the system
may be chaotic even when the geometry yields an otherwise fully
integrable one-particle case, as we shall see below.

On the other hand, recent experimental realizations of billiards, such
as suitably shaped resonators and quantum dots \cite{Stock,Marcus},
have allowed the study of the quantum manifestations of well-known
classical non-integrability in some billiards
\cite{Bunimovich,McDonald}. In the case of quantum dots, disagreement
between theory and experiment has been attributed to geometrical
factors \cite{Marcus}. A considerable amount of theoretical work
exists on the effect that geometry has on the integrability of
dynamics in billiards \cite{Bunimovich,Sinai,Kozlov,Galperin}, as well
as on their quantum analogs \cite{McDonald,Primack}. However, the
possibility of more than one particle in the quantum dot leaves the
usual one-particle approach incomplete.  In fact, some experiments
have pointed out the importance of electron-electron interaction on
various features observed in such mesoscopic systems \cite{Sivan}.  In
this article, we explore the role of the electrostatic interaction
introduced when two particles are in the billiard. A formalism for
billiards in any dimensions is developed, and as an example, we apply
it to the one-dimensional case.  Since we are interested in the role
of the electrostatic interaction in mesoscopic systems, we consider
particles interacting through a screened Coulomb potential.

 {\em The hyperbilliard}.~  The problem of two point masses moving along a finite line and
suffering elastic impacts with the end walls and between themselves,
can be transformed to the motion of one {\it `particle'} moving in a
triangular billiard. The coordinates of the {\it particle} in this
billiard are the coordinates of the original masses.  The ratio of the
masses determines the integrability of the system \cite{Sinai}, being
regularizable for a  particle mass ratio of 1 and 3 (or
$\frac{1}{3}$) \cite{Kozlov}. 

 We now introduce an interaction between the particles and consider the
$d$-dimensional case. Let ${\bf q}_i,{\bf p}_i$, $(i=1,2)$ be the
position and linear momentum of the $i$-th particle.  The motion takes
place in a $d$-dimensional billiard, a compact simply-connected region
of $R^d$ whose boundary is denoted by  $\Gamma$. We assume that
$\Gamma$ is piecewise smooth and defined by $\nu$ surfaces,
$\Gamma_j=\{ {\bf q}:f_j({\bf q},\alpha_j)= 0\}, j=1,...,\nu$, where
$f_j$ and $\alpha_j$ denote the function and the set of constants which
characterize the $j$-th surface. These functions define subspaces of
dimension $d-1$ in $R^d$. To fix ideas, we restrict ourselves to flat
surfaces, {\em i.e}., for $d=2$ (3) the billiards are simple polygons
(polyhedrons).

 The formalism developed here can be applied to any central-force
interaction between the particles. We have selected the  screened
Coulomb potential, {\it i.e.}, the Yukawa potential given by
 $ V( {\bf q_1}, {\bf q_2}) = e^{ -\lambda |{\bf q}_2 - {\bf q}_1 |} /
 |{\bf q}_2 - {\bf q}_1|$,
 where $\lambda^{-1}$ is the screening length. Notice that this
potential goes to a $\delta-$function when $\lambda \rightarrow
\infty$.
 In this limit, the particles behave as bouncing hard-core balls, {\it
i.e.}, non-interacting impenetrable point particles, for which the
dynamics can be integrable, as described above.  Hence, for a given
energy,  $\lambda$ plays the role of the perturbation parameter. Due to
the interaction, a finite value of $\lambda$ determines the finite
effective radius of the particles for a given total energy, as
described below.

Considering for simplicity identical-mass particles ($m_1 = m_2 =1$),
the Hamiltonian for the system is written as 
 \begin{eqnarray} 
  H = \frac{p_1^2}{2} +
 \frac{p_2^2}{2} + V({{\bf q}_1,{\bf q}_2}) +
 \sum_{i=1}^2 \sum_{j=1}^{\nu} U[f_j({\bf q}_i,\alpha_j)],
 \end{eqnarray} 
 where the function $U[f_j({\bf q}_i,\alpha_j)]$ represents the
infinite repulsion potential exerted by  the $j$-th hard wall on the
$i$-th particle.  Analytically, this function could be written in terms
of Heaviside functions with a large prefactor. In practice, the  normal
component of the velocity of the incident particle will be reversed at
the moment of bouncing on the billiard walls.  

The Hamilton equations can be written as $\dot{\bf q}_i  = {\bf p}_i$,
and
 $ 
 \dot{\bf p}_i  = -\nabla_{q_i} V({\bf q}_1,{\bf q}_2) +
\sum _{j=1} ^{\nu}{\bf A}_j({\bf p}_i) \delta[f_j({{\bf q}_i},\alpha_j)] ,
 \label{eqpg}
 $ 
 where $i=1,2$ and the vector function ${\bf A}_j$ represents the
change of momentum due to the bounce on the $j$-th wall.  Given $d$,
the number of degrees of freedom is $2d$. Hence, the phase space of the
system is $4d$-dimensional.

We now introduce a transformation to center-of-mass and relative
coordinates ${\bf R}  =  ({\bf q}_1 + {\bf q}_2)/M  $, and ${\bf r}  = 
{\bf q}_2 - {\bf q}_1$, respectively, where the total mass $M=2$, and
the reduced mass $\mu = \frac{1}{2}$. These equations define a new space of
coordinates ${\bf \rho} = ({\bf r,R})$, which is $2d$-dimensional. In
this space, we have a new set of equations for the boundary of the
billiard, say $F_j({\bf \rho},\alpha_j)$, $j=1,...,\nu$. Every function
$F_j$ now defines a subspace of $2d-1$ dimensions in $\rho$-space.
 
The Hamilton equations are transformed then to $\dot{\bf r}  =  {\bf p}/\mu 
$, $\dot{\bf R}  =   {\bf P}/M $, and
 \begin{eqnarray}
 \dot{\bf p} & = & -\nabla_{r} V(r) +  
            \sum _{j=1} ^{\nu}{\bf A}_j ({\bf p,P}) \, \delta[F_j({\bf
r,R},\alpha_j)] \, ,   \nonumber \\
 \dot{\bf P} & = & \sum _{j=1} ^{\nu}{\bf B}_j({\bf p,P}) \, \delta[F_j({\bf
r,R},\alpha_j)]. \label{eqpr2}
 \end{eqnarray}
 As before, ${\bf A}_j$ and ${\bf B}_j$ represent the change of the
momenta {\bf p} and {\bf P}, respectively, due to the bounce on the
$j$-th wall.

Notice that these equations describe the motion of {\it one particle}
in  the $\rho$ hyperspace, {\it i.e.}, we have constructed the {\em
hyperbilliard}.  The description of a system composed by a few masses
in terms of one $particle$ in a hyperspace has been used for several
cases, including billiards \cite{Sinai,Kozlov,Leinass}.
 Usually, the hyperspace is constructed without introducing
transformations of the coordinates. Here, however, the change to the CM
coordinates allows one to get a map in a simple way. 

Notice that bounces of the $particle$ in the hyperbilliard correspond
to bounces of the masses in the real billiard.    The walls of the
billiard cause the breaking  of the translational symmetry of the
system, and as a consequence, the CM momentum is no longer a constant
of motion. In the case of non-interacting and equal-mass particles, the
changes in the CM momentum are determined only by the geometry of the
billiard.  In our case, however, the interaction couples the CM and
relative momenta after a bounce, which in turn depend on the momenta of
each of the original masses.  The rotational symmetry is also broken in
general and the generator of rotations is no longer a constant of
motion either.

 {\em The map}.~ 
Hamilton equations in $\rho$-space indicate that {\em between} bounces
the {\em particle} moves freely along the CM coordinate whereas the
central force $V(r)$ acts only along ${\bf r}$. The motions are
independent, and only become correlated at each bounce, as the
corresponding momenta are changed while keeping the total energy
constant.  We take advantage of this fact: Consider that the {\em
particle} at the $n$-th bounce has the coordinate $ {\bf \rho} _n = (
{\bf r} _n, {\bf R} _n )$. The condition that the time spent by the
{\it particle} until the next bounce on the $j$-th wall be the same
along the ${\bf r}$ and ${\bf R}$ coordinates,
 \begin{equation} 
 \tau_r (  \rho_n , \rho_{n+1}  ) = \tau_{R_k} (\rho_n , 
 \rho_{n+1}  ), \hspace{1cm} k=1,...,d, 
 \label{tau}
 \end{equation} 
 represents an interesting opportunity. Here, $ \tau _r$ ($ \tau
_{R_k}$) refers to the time along the relative ($k$-component of CM)
coordinate. The times $\tau _{R_k}$ for the free motion between
collisions can be calculated easily. The l.h.s.\ in (\ref{tau}) can be
obtained by noting that the motion along ${\bf r}$ becomes separable
and the time $\tau_r$ can then be calculated by quadratures, as
illustrated below.  Equations (\ref{tau}) and the equation corresponding
to $F_j$
 result in a set of nonlinear algebraic equations for $\rho_{n+1}$.  We
call this set the $map$ of the billiard since it indeed expresses
$\rho_{n+1}$ in terms of $\rho_{n}$. This procedure can be easily
carried out at least formally in the general case.  Notice that this
map has not been obtained by means of the usual linearization
procedure \cite{Lichtenberg}, but rather as an extension of Benettin's
procedure \cite{Benettin2}. The 1D case, explained in detail now,
provides a clear example of this procedure.

 {\em The 1D billiard}.~
  This system is defined by walls at the end points $q = \pm \frac{1}{2}$ .
Because of the interparticle repulsion, the particle 1 (2) never
reaches the boundary 2 (1). This implies that $ A_j$, here associated
with bounces {\em on} the $j$-th wall, will describe bounces of the
$j$-th particle only. The Hamilton equations for the $r-R$ coordinates
are then $p  =  \mu \dot{r}$, $ P = M \dot{R}$, and
 \begin{eqnarray}
 \dot{p} & = &- \frac{d V(r)}{dr} + A_1 \,
 \delta(R+\frac{1}{2}-\frac{r}{2}) + A_2 \, \delta(R-\frac{1}{2}- \frac{r}{2}),
 \nonumber \\
 \dot{P} & = & B_1 \, \delta ( R+\frac{1}{2}-\frac{r}{2}) + B_2 \,
 \delta(R-\frac{1}{2}- \frac{r}{2}).
 \end{eqnarray} 
 According to the arguments of the $\delta$-functions, the point
boundaries are transformed into lines in $\rho$-space, which define a
billiard with an isosceles-triangle shape, similar to the case of
non-interacting hard core particles \cite{Sinai}, although here is in
the $r-R$ space. The base of this triangle in our case acts as a
repulsive wall of potential $V(r)$.  The closest approach to the
repulsive wall by the {\em particle} (the turning point), depends on
the energy associated with the relative motion, $\epsilon = E -P^2/2M$.

 The functions giving the change of momentum are simple. For example,
for bounces of the $i-th$ particle (on $i-th$ wall) we have $ A_i = -p
\pm 2 \mu P/ M$, where $P$ and $p$ are the momenta {\em before} the
collision, and the $+$ ($-$) sign refers to $i=1$ (2).

For a pure Coulomb potential ($\lambda = 0$), $\tau_r$ can be
calculated analytically, so Eq.\ (\ref{tau}) can be written  in the
form
 \begin{equation}
 \tau \{ T _\epsilon (\rho _n), T _\epsilon (\rho _{n+1})\} = \left| 
 \frac{R _{n+1} - R _n}{P/M} \right|,
 \label{map}
 \end{equation}
 where $\tau$ is the time elapsed going from $\rho _n$ to $\rho_{n+1}$,
expressed in terms of the time $T$ spent by the particle from the
turning point to $\rho$ 
 \begin{equation} 
 T _\epsilon (\rho) = \frac{ rp}{2 \epsilon} +
\left(\frac {\mu}{2 \epsilon ^3} \right)^{\frac{1}{2}} \cosh^{-1} (r
\epsilon) ^{\frac{1}{2}}. \\
 \end{equation}
 For $\lambda \leq 1$ we can expand $V(r)$ to first order and obtain
the same expression, except that $E$ is shifted to $E-\lambda$.  For
all different initial conditions there are only a few possible
trajectories which can be determined by analyzing the momenta.  A
simple algorithm can then be obtained to determine the Poincar\'e
surfaces of section. (The details of the motion in $\rho$-space will
be presented elsewhere.) This nontrivial algebraic map provides a
full description of the dynamics. Its use simplifies calculations a
great deal, and allows one to better characterize the system, as we
describe below.

To characterize the dynamics, we determine the Poincar\'e section (PS)
at a phase such that one of the masses is fixed, say, as it just
bounces on the wall.  Then we plot the position and momentum of the
other mass.  Because of the indistinguishability of the particles, the
topology of the PS does not depend on which mass is selected. In fact,
the surfaces are identical, except for left-right exchange symmetry.

Between bounces on the walls, the masses approach each other a
distance given by their relative energy.  The shortest distance of
approach $r_m$, can be considered as twice the minimum effective
radius of the particles (for zero CM energy). This effective radius is
a characteristic of the system and its dependence on the total energy
$E$ and the inverse screening length $\lambda$ is shown in
Fig. \ref{rm}, as obtained from the condition $E = V(r_m)$. When
$\lambda \gg 1$, the interaction is short-ranged, which results in
nearly-free particles for some moments.  Note that for all values of
$E$ and $\lambda$, the initial condition $q_1^{(0)}= -\frac{1}{2},
q_2^{(0)} =\frac{1}{2},p_1^{(0)}= \surd [E-V(1)], p_2^{(0)}=
-p_1^{(0)}$, corresponds to a periodic orbit and we call it the {\it
symmetric} motion.

We now fix $E=1.56$ and change $\lambda$. Fig.\ \ref{fq1}a shows the
PS for the ``short-range potential'' case, $\lambda =20$. These
results were obtained by direct numerical integration of the equations
of motion. Here $q_1=-\frac{1}{2}$, which means that we are plotting
the position and momentum of the particle 2 as particle 1 is at its
(left) edge of the billiard. According to the KAM theorem
\cite{Gutzwiller}, some invariant tori will be preserved under the
interaction, although they are somewhat deformed. These periodic or
quasiperiodic orbits lie inside the {\it primary islands of
stability}, which in our case are situated around the fixed point
corresponding to the symmetric motion, the latter one represented by
means of $\times$.  Higher values of $\lambda$ present a similar PS,
but the chaotic region fills more and more of the available space
consistent with $E$ fixed.  It is clear that only for the case of
point particles, {\it i.e.} zero screening length ($1/\lambda=0$), the
bouncing-ball behavior is observed.  It is also possible that the
infinite energy limit (with $r_m=0$) would be similar, as Fig.\ 1 also
suggests.

Secondary islands appear for $\lambda\approx 1$. These islands are due
to the inter-particle interaction \cite{Lichtenberg} and correspond to
correlated motion, as when for example, particle 1 bounces twice and
the other once.  For each pair of values ($\lambda$, $E$) there is a
specific island structure. The orbits in the secondary islands become
unstable for the ``short-range'' case because for some instants the
particles are nearly free, the memory of the previous motion is lost,
and the correlation is destroyed.  Hence, as $\lambda$ decreases
(Fig.\ \ref{fq1}b), the number of stability islands increases.  In
this case $q_2=\frac{1}{2}$, so that the PS shows the position and
momentum of particle 1 when particle 2 is at the right edge of the
billiard.  Notice that the available region of space decreases for
smaller $\lambda$, as the total energy is fixed and the inter-particle
potential energy has a stronger confining effect.

The results using the $map$ described before are now presented. Figure
\ref{l0} shows the PS for $\lambda = 0$ obtained by solving ($a$) the
differential Hamilton equations in $q$-space, and ($b$) the algebraic
Eq.\ (\ref{map}). The graph obtained by means of the latter has been
reflected about $q_2=\frac{1}{2}$ for easy comparison.  The two
$\lambda =0$ PS are topologically identical (if traversed in different
sequences). The agreement is excellent even for $\lambda \approx 1$,
while the computation time is substantially reduced ($\sim 10^3$
times) if the map is used.

 Using the map, we have calculated the Lyapunov exponent $\sigma$ for
$\lambda \leq 1$, following the procedure of Ref.\ \cite{Benettin2}.
Figure \ref{fq1} shows that, as $\lambda$ increases, the fraction of
phase space filled by the chaotic sea increases also.  This is
reflected in the Lyapunov exponent (not shown), which for a constant
energy ($E=1.56$) increases monotonically (from 0.34 to 0.59) with the
inverse screening length $\lambda$ (0 to 1).  Increasing energy
produces similar curves with ever larger values of $\sigma$.

A general formalism for two interacting particles in a $d$-dimensional
billiard has been presented. The one-dimensional case with a screened
Coulomb potential was shown to exhibit {\it soft chaos}.  Only in the
case of infinite screening length (or energy), the particles can be
considered as bouncing balls. These results suggest that the effects
of electrostatic interaction between electrons in quantum dots, for
example, may play a very important role in the quantum-classical
correspondence and they should be considered when these systems are
studied. The analysis of the quantum mechanical analog of the billiard
system described in this work is now in progress.

We thank G. Luna for valuable suggestions. Partial financial support
by CONACyT, M\'exico and US DOE grant no.\ DE--F02--91ER45334 is
acknowledged.  SEU acknowledges support of the A. v.\ Humboldt
Foundation.



\begin{figure}
 \caption{The closest approach between the particles, $r_m$, as a
function of the inverse screening length $\lambda$, for different
energies $E$.} \label{rm}
 \end{figure}
 
 \begin{figure} 
 \caption{Poincar\'e sections for $E=1.56$. a)
$q_1=-\frac{1}{2}$, $\lambda= 20$, b) $q_2=\frac{1}{2}$, $\lambda=0.6$. Crosses
($\times$) indicate the symmetric periodic motion.} \label{fq1}
 \end{figure}

 \begin{figure}
 \caption{Poincar\'e sections for the selected energy $E=1.56$,
$\lambda$=0 and $q_1=-\frac{1}{2}$. a) Solving the Hamilton equations and b)
using the $map$.} \label{l0} 
 \end{figure} 

 \end{document}